\def\year{2019}\relax
\documentclass[letterpaper]{article} 
\usepackage{aaai19}  
\usepackage{times}  
\usepackage{helvet}  
\usepackage{courier}  
\usepackage{url}  
\usepackage{graphicx}  
\usepackage{capt-of}
\expandafter\def\expandafter\UrlBreaks\expandafter{\UrlBreaks
  \do\a\do\b\do\c\do\d\do\e\do\f\do\g\do\h\do\i\do\j%
  \do\k\do\l\do\m\do\n\do\o\do\p\do\q\do\r\do\s\do\t%
  \do\u\do\v\do\w\do\x\do\y\do\z\do\A\do\B\do\C\do\D%
  \do\E\do\F\do\G\do\H\do\I\do\J\do\K\do\L\do\M\do\N%
  \do\O\do\P\do\Q\do\R\do\S\do\T\do\U\do\V\do\W\do\X%
  \do\Y\do\Z}
\usepackage{floatrow}
\newfloatcommand{capbtabbox}{table}[][\FBwidth]
\usepackage{multirow}
\usepackage[inline]{enumitem}
\DeclareFloatFont{small}{\small}
\usepackage[dvipsnames]{xcolor}
\definecolor{Orange}{HTML}{EDA486}
\definecolor{Yellow}{HTML}{F5FF8A}
\definecolor{Green}{HTML}{7FE947}
\definecolor{Magenta}{HTML}{C29DC8}
\definecolor{Cyan}{HTML}{6FDCDB}

\usepackage{booktabs}

\usepackage{tcolorbox}

\newtcbox{\xmybox}[1]{
    on line, arc=3pt, colback=#1,
    before upper={\rule[-3pt]{0pt}{10pt}}, boxrule=0pt, colframe=white,
    boxsep=0pt, left=3pt, right=3pt, top=1.5pt, bottom=0pt
}
\renewcommand{\colorbox}[2]{\xmybox{#1}{#2}}

\usepackage{array} 		
\newcolumntype{L}[1]{>{\raggedright\let\newline\\\arraybackslash\hspace{0pt}}p{#1}}
\newcolumntype{C}[1]{>{\centering\let\newline\\\arraybackslash\hspace{0pt}}p{#1}}
\newcolumntype{R}[1]{>{\raggedleft\let\newline\\\arraybackslash\hspace{0pt}}p{#1}}

\begin{document}

\title{Different Spirals of Sameness: A Study of Content Sharing in Mainstream and Alternative Media}

\author{Benjamin D. Horne\textsuperscript{*}, Jeppe N{\o}rregaard\textsuperscript{\dag}, and Sibel Adal{\i}\textsuperscript{*} \\
Rensselaer Polytechnic Institute\textsuperscript{*}, Technical University of Denmark\textsuperscript{\dag}\\
horneb@rpi.edu, jepno@dtu.dk, adalis@rpi.edu
}

\maketitle
\begin{abstract}
In this paper, we analyze content sharing between news sources in the alternative and mainstream media using a dataset of 713K articles and 194 sources. We find that content sharing happens in tightly formed communities, and these communities represent relatively homogeneous portions of the media landscape. Through a mix-method analysis, we find several primary content sharing behaviors. First, we find that the vast majority of shared articles are only shared with similar news sources (i.e. same community). Second, we find that despite these echo-chambers of sharing, specific sources, such as The Drudge Report, mix content from both mainstream and conspiracy communities. Third, we show that while these differing communities do not always share news articles, they do report on the same events, but often with competing and counter-narratives. Overall, we find that the news is homogeneous within communities and diverse in between, creating different spirals of sameness. 
\end{abstract}

\section{Introduction} 
Researchers in Communications have studied content sharing in journalism for quite some time~\cite{boczkowski2010news,graber1971press,noelle1987theevent,shoemaker2013mediating}. This long line of research has shown that news organizations often imitate each other in order to be competitive and meet demand. Various reasons for this behavior have been discussed, such as the popularity of the Internet~\cite{mitchelstein2010online} and changes in the news demand structure~\cite{boczkowski2010news}. This behavior was even discussed as early as the 1950's, where it was said that ``many newspapers feature the same news stories atop their front pages~\cite{boczkowski2010news,breed1955newspaper}.'' It has been argued that because of this increased content copying, the news has become homogeneous and significantly less diverse~\cite{boczkowski2010news,klinenberg2005convergence,glasser1992professionalism}.  

However, today this homogenized view of the news has been complicated by the rise of ``alternative'' media. Specifically, the rise of false, hyper-partisan, and propagandist news producers has created a media landscape where there are competing narratives around the same event~\cite{starbird2017examining} and no gatekeepers to curate quality information~\cite{reese2009journalists,allcott2017social,mele2017combating}. Thus, we may have a more diverse set of news to read than in years past, but the standards of quality have wavered, creating a new set of concerns. 

This rise in low-quality and potentially malicious news producers has been the focus of many recent studies such as those focusing on detecting false content~\cite{potthast2017stylometric,popat2016credibility,singhania20173han,horne2018accessing,baly2018predicting}. Some other studies have focused on the tactics used to spread low-quality news, such as the use of social bots~\cite{shao2017spread} and the structures of headlines to get higher attention and clicks~\cite{horne2017just,chakraborty2016stop}. One lesser studied area of the alternative media universe is content sharing (or content republishing, imitation, replication). While in the past content sharing in the mainstream media was used to meet demand, be timely, and keep up with competing news agencies, it may be used more maliciously in today's news environment. For example, just as bot-driven misinformation in social networks, content sharing can be used to make particular stories or narratives seem more important, more widely reported, and thus, more credible. 

In this paper, we begin to explore this behavior. Specifically, we analyze content sharing on a large dataset (713K articles and 194 sources) across both the mainstream and alternative landscapes, with news sources of varying veracity. We show that, when formulated as a network, news producers share content in tightly connected communities. Furthermore, these communities represent distinct parts of the media ecosystem, such as U.S. mainstream media, left-wing blogs, and right-wing conspiracy media. With this community framework, we employ mix-methods analysis to better understand what types of content sharing behavior exist within and between these communities. We observe four primary practices in this data. First, news content is often replicated in echo-chambers, where the copied content is only published by other producers within the community. This may mean a high quality investigative piece of reporting or a wild conspiracy theory may be equally copied within the network that originated it. Second, despite the tight community structure of the content sharing network, specific sources mix content from both mainstream news and conspiracy news. This behavior illustrates a dangerous practice, which can falsely elevate the perceived credibility of conspiracy-spreading sources. Third, many news articles are not shared across communities, but the broad topics and events featured in the articles can be very similar, many times in the form of competing contemporaneous narratives. Lastly, we observe a more unique behavior in which the conspiracy media reacts to a mistake in the mainstream media, ultimately providing a reason to distrust the mainstream media. 

Overall, we find that the homogeneous view of news (or to borrow a term from \cite{boczkowski2010news}: ``spiral of sameness'') still exist, but those ``spirals of sameness'' are, for the most part, different in each distinct parts of the news ecosystem. 
Within the same community, multiple processes work simultaneously to
amplify certain narratives around current events as well as to undermine the credibility of some high quality news outlets. In essence, this "spiral of sameness"
now also actively works to create a type of otherness that feeds the creation of more divisive news and an overall confusing information environment.

\section{Related Work}
There are two recent studies that have focused on content sharing in today's media ecosystem. The first study on content sharing in alternative media focuses on a specific topic in 2016: the Syrian Civil Defense~\cite{starbird2018ecosystem}. This study uses mix-methods to analyze the content replication practices by alternative news sites reporting on various aspects of the Syrian Civil Defense. The authors used Twitter as a the starting point of the data collection, and extended to the websites cited in the Twitter data. With this data, the authors demonstrated the spread of competing narratives through content sharing. They found that the alternative news sources had both news-wire as well as news aggregator type services. Additionally, they found that a small number of authors generate content that is spread widely in the alternative news. They also found that government-funded media were prevalent in the production these anti-White Helmet narratives. 

The second study approaches content republishing from a more general setting~\cite{horne2018exploration}. Specifically, the authors collected news data from 92 news sources, that included both mainstream and alternative news. The articles collected were not focused on any specific topic as was done in the study discussed above~\cite{starbird2017examining}. Horne and Adal{\i} collected this data live from each news source, and thus, were able to gather timestamps with each article. With this data, they created directed networks of news sources, where each edge represents some number of nearly identical articles. They found that despite many articles being copied verbatim, the headlines of the articles often changed. These headline changes differed between the alternative media and the mainstream media, where the alternative media often changed emotional tone and the mainstream media often change structural features. Furthermore, the authors found that most alternative content is written by very few authors, just as was found in~\cite{starbird2017examining}. 

In contrast to the two previous works, our work uses a much larger dataset that covers a long period of time and a large number of topics/events. Additionally, our analysis incorporates both exact and partial matching algorithms, providing a more extended look at content sharing than the previous two studies. Lastly, we utilize external credibility and bias assessments to better characterize the sources who are sharing content, which allows us to conduct extensive new case studies that have not been shown in the literature. We hope that this work, in combination with these previous works, can be a strong building-block in developing theory about content sharing as a disinformation tactic.  
\section{Methods}
\subsection{Data}
We collected articles from a broad spectrum of sources. We scraped the RSS feeds of each news source twice a day starting on 02/02/2018 using the Python
libraries feedparser and goose. For source selection, we start with mainstream outlets (from both the U.S. and the U.K.) and alternative sources that are mentioned in other misinformation studies~\cite{starbird2017examining,horne2018exploration,baly2018predicting}. We then use the Google Search API to expand the number of sources in the collection. Specifically, we query Google with the titles of the previously collected articles and add any source that appears in the top 10 pages of Google and is not already in our collection list. This process is repeated until we have a large sample of sources from both mainstream and alternative news. In addition to scraping article content, we capture the UTC timestamp of when the article was published. Note, we do not include small local news sources or sources that did not have operational RSS feeds, which significantly reduces the size of the expected source set. Our final dataset contains 194 sources with over 713K articles between 02/02/2018 and 11/30/2018. Since this collection process happens multiple times a day, we have nearly every article published by a source after it is added to the collection. 

\subsection{Building Content Sharing Networks}~\label{networkbuilding}
Once our data collection is complete, we construct a verbatim content sharing network. We take a similar, but more refined, approach to \cite{horne2018exploration}.

We employ a three step method to build the network:
\begin{enumerate}
    \item We build a Term Frequency Inverse Document Frequency (TFIDF) matrix for each 5 day period in the dataset. For each pair of article vectors, we compute the cosine similarity between them. Following the same process in~\cite{starbird2018ecosystem} and~\cite{horne2018exploration}, we choose article pairs with cosine similarity of 0.85 or above. These extracted article pairs are nearly identical, excluding potentially different interpretations of the same story. The 5 day window is used for computational reasons, to reduce the size of pairwise comparison matrix.
    \item For each pair, we order them by the UTC timestamp, as to create directed edges from the original article to the copied article. 
    \item Each article that is a copy, can only copy from one original article, but an article being copied can be copied by multiple other news sources. Thus, if multiple pairs were extracted (i.e. 4 verbatim articles would create 12 unordered pairs, 6 ordered pairs after timestamp ordering), we match a copying article to an older article with highest cosine similarity. If there are ties, we pick the oldest article as the original article. 
\end{enumerate}

After this process, we perform some manual verification of pairs, as some UTC timestamps are slightly off due to a news source updating or republishing an article. Once we are confident in the article pairs, we build a directed network, where nodes are news sources and edges are articles copied between the sources. Each edge is weighted by the number of articles copied and is directed from original source to the source that copied. We find that 160 sources out of the 194 sources in the dataset copied an article or had an article copied from them at least once during the 10 month period.

\subsection{Finding Communities}
Next, we use the modularity maximization algorithm designed specifically for directed networks~\cite{leicht2008a} to determine communities in the network. This algorithm uses simple network statistics to compute probabilities of edges between a set of nodes. The modularity score is a measurement of how improbable the distribution of edges within a set of communities are, compared to the distribution based on the simple statistics. By maximizing the modularity, we find communities which have a surprisingly high number of internal edges when compared to the expectation. The python implementation we used\footnote{\url{zhiyzuo.github.io/python-modularity-maximization/}} determines both the number of communities as well as the communities themselves. The network can be found in Figure~\ref{nets1} and is colored with the detected communities.

\begin{table}[h!]
    \definecolor{newguardred}{RGB}{201,32,39}
    \definecolor{newguardgreen}{RGB}{66,177,73}
    \definecolor{newguardgrey}{RGB}{171,170,171}
    \renewcommand{\arraystretch}{1.3}
    \centering

    \begin{tabular}{cccc}
    \toprule
    {} & \colorbox{newguardgreen}{\color{white} \textbf{Credible}} &       \colorbox{newguardred}{\color{white} \textbf{Not Credible}} & \colorbox{newguardgrey}{\color{white} \textbf{Unknown}} \\
    \midrule
    \textbf{\colorbox{Orange}{ O }} &    3 \hspace{1ex}(8\%) &   2 \hspace{1ex}(5\%) &  34 \hspace{1ex}(87\%) \\
    \textbf{\colorbox{Yellow}{ Y }} &   9 \hspace{1ex}(27\%) &  5 \hspace{1ex}(15\%) &  19 \hspace{1ex}(57\%) \\
    \textbf{\colorbox{Green}{ G }} &  26 \hspace{1ex}(74\%) &   2 \hspace{1ex}(6\%) &   7 \hspace{1ex}(20\%) \\
    \textbf{\colorbox{Magenta}{ M }} &   8 \hspace{1ex}(47\%) &   0 \hspace{1ex}(0\%) &   9 \hspace{1ex}(53\%) \\
    \textbf{\colorbox{Cyan}{ C }} &   5 \hspace{1ex}(14\%) &   0 \hspace{1ex}(0\%) &  30 \hspace{1ex}(86\%) \\
    \bottomrule
    \end{tabular}

    \caption{Number of sources in each community within the three main labels of NewsGuard. Unknown means the source has not been labeled by NewsGuard, this does not necessarily mean they are unreliable sources.}
    \label{tbl:newsguard_labels}
\end{table}
\begin{table}[h!]
    \centering
    \renewcommand{\arraystretch}{1.3}

    \begin{tabular}{ccccc}
    \toprule
    {} &       \includegraphics[height=2ex]{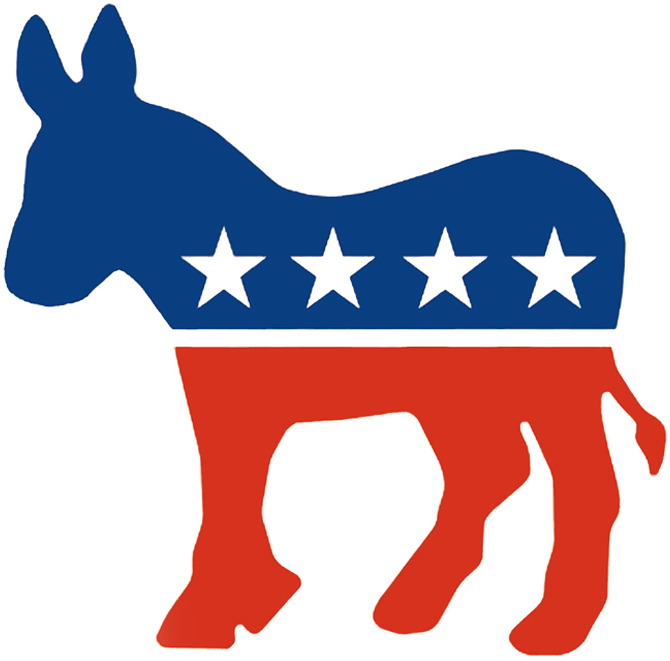} &    \includegraphics[height=2ex]{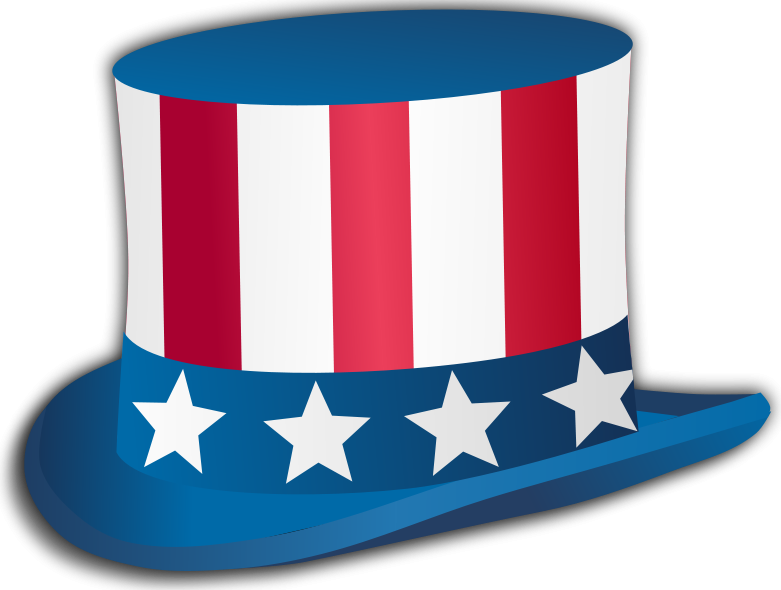} &      \includegraphics[height=2ex]{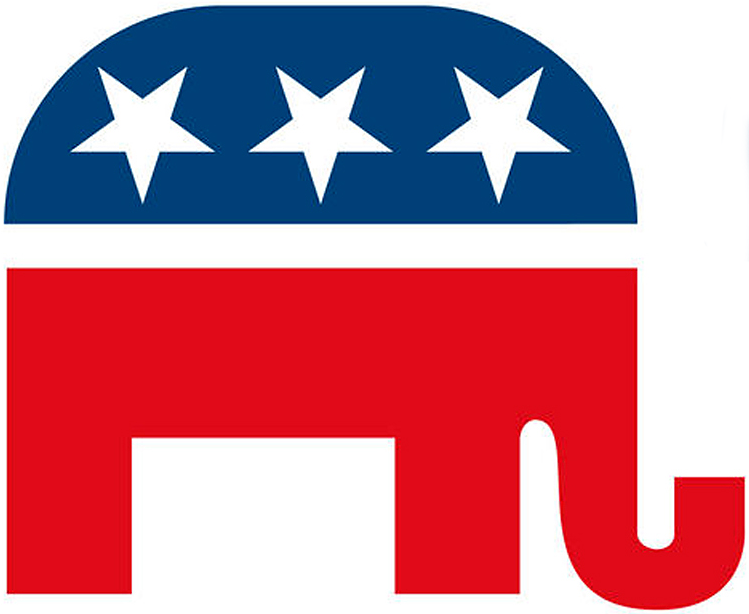} &    \includegraphics[height=2ex]{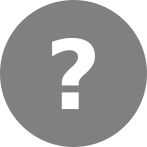} \\
    \midrule
    \textbf{\colorbox{Orange}{ O }} &   6 \hspace{1ex}(15\%) &   5 \hspace{1ex}(13\%) &   4 \hspace{1ex}(10\%) &  24 \hspace{1ex}(62\%) \\
    \textbf{\colorbox{Yellow}{ Y }} &   4 \hspace{1ex}(12\%) &    1 \hspace{1ex}(3\%) &  23 \hspace{1ex}(70\%) &   5 \hspace{1ex}(15\%) \\
    \textbf{\colorbox{Green}{ G }} &  14 \hspace{1ex}(40\%) &  12 \hspace{1ex}(34\%) &   5 \hspace{1ex}(14\%) &   4 \hspace{1ex}(11\%) \\
    \textbf{\colorbox{Magenta}{ M }} &  10 \hspace{1ex}(59\%) &    1 \hspace{1ex}(6\%) &    0 \hspace{1ex}(0\%) &   6 \hspace{1ex}(35\%) \\
    \textbf{\colorbox{Cyan}{ C }} &   7 \hspace{1ex}(20\%) &  11 \hspace{1ex}(31\%) &   4 \hspace{1ex}(11\%) &  13 \hspace{1ex}(37\%) \\
    \bottomrule
    \end{tabular}

    \caption{Counts of political leaning of sources in each community, based on data from Media Bias / Fact Check. The leanings are \includegraphics[height=1.5ex]{Figures/leaning/donkey.png} left, \includegraphics[height=1.5ex]{Figures/leaning/hat.png} center, \includegraphics[height=1.5ex]{Figures/leaning/elephant.png} right and \includegraphics[height=1.5ex]{Figures/leaning/unknown.png} unknown.}
    \label{tab:community_leaning}
\end{table}
\begin{table}[h]
    \centering
    \renewcommand{\arraystretch}{1.3}
    \begin{tabular}{cccccc}
    \toprule
    {} &    \frame{\includegraphics[height=2ex]{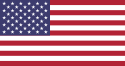}} &          \frame{\includegraphics[height=2ex]{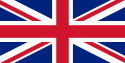}} &  \frame{\includegraphics[height=2ex]{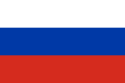}} &        \frame{\includegraphics[height=2ex]{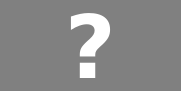}} &    \includegraphics[height=2.4ex]{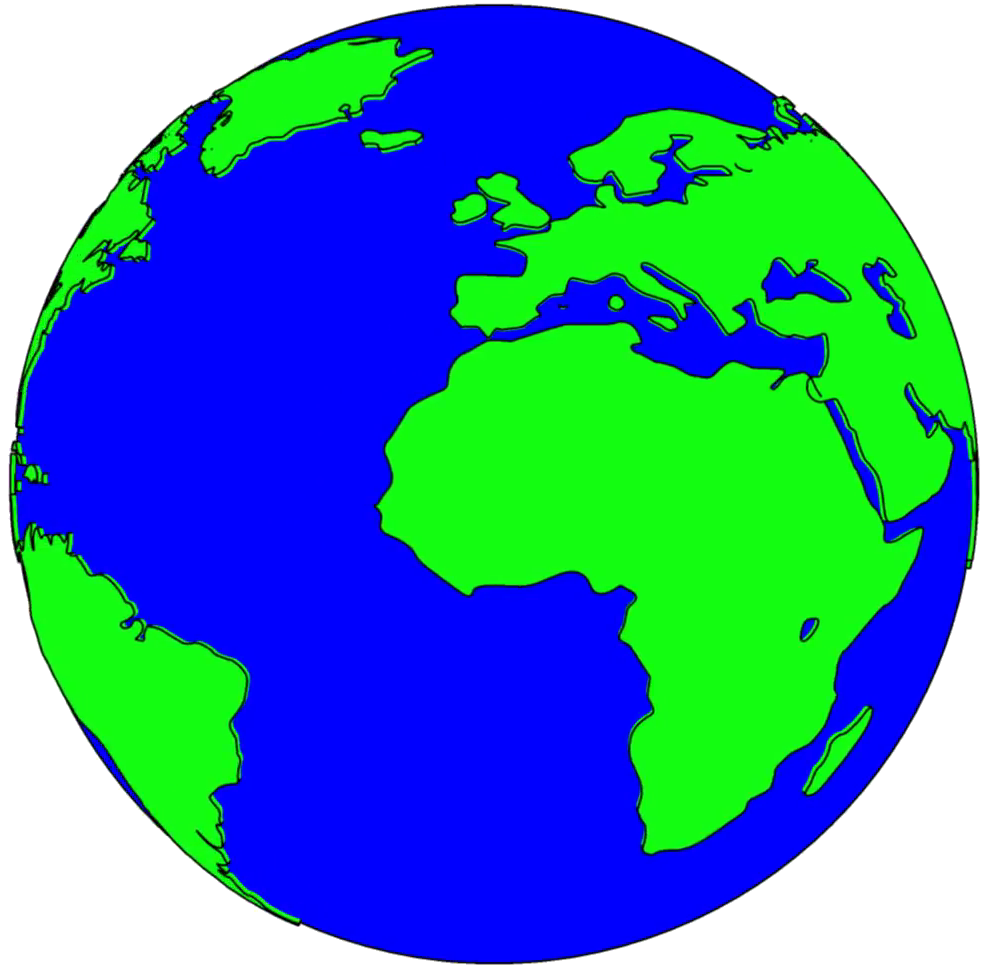} \\
    \midrule
    \textbf{\colorbox{Orange}{ O }} &  21 (54\%) &    0 (0\%) &  8 (21\%) &  7 (18\%) &  3 (8\%) \\
    \textbf{\colorbox{Yellow}{ Y }} &  31 (94\%) &    0 (0\%) &   0 (0\%) &   2 (6\%) &  0 (0\%) \\
    \textbf{\colorbox{Green}{ G }} &  31 (89\%) &    2 (6\%) &   0 (0\%) &   0 (0\%) &  2 (6\%) \\
    \textbf{\colorbox{Magenta}{ M }} &  16 (94\%) &    1 (6\%) &   0 (0\%) &   0 (0\%) &  0 (0\%) \\
    \textbf{\colorbox{Cyan}{ C }} &   7 (20\%) &  22 (63\%) &   2 (6\%) &   1 (3\%) &  3 (9\%) \\
    \bottomrule
    \end{tabular}

    \caption{Counts of countries of origin for sources in each community. The countries are \frame{\includegraphics[height=1.4ex]{Figures/flags/usa.png}} USA, \frame{\includegraphics[height=1.4ex]{Figures/flags/russia.png}} Russia, \frame{\includegraphics[height=1.4ex]{Figures/flags/uk.png}} United Kingdom, \frame{\includegraphics[height=1.4ex]{Figures/flags/unknown.png}} unknown country and \includegraphics[height=1.7ex]{Figures/flags/globe.png} other countries (Canada, Cyprus, France, Germany, New Zealand, Qatar, Ukraine or Venezuela)}
    \label{tab:community_countries}
\end{table}

\subsection{Characterizing Communities}
In order to better understand what types of sources exist in each community, we utilize source-level analyses done by several platforms: NewsGuard\footnote{\url{newsguardtech.com}}, Media Bias Fact Check\footnote{\url{mediabiasfactcheck.com}} (MBFC), Allsides\footnote{\url{allsides.com}}, and an article published by BuzzFeed\footnote{\url{https://bit.ly/2OoYztU}}. 
NewsGuard uses a large team of trained journalists to review news outlets, in order to inform readers about the sites, as well as organizations who work with or publish ads on the news sites. NewsGuard assesses nine journalistic criteria which are combined into a green (good) or a red label (bad).
MBFC is a platform that analyzes news sources to determine their credibility using trained team. We combine their factual-reporting score with NewsGuard's credibility label, for a final label of source reliability. We aggregate by normalizing both scores from -1 to 1 and adding them. Sources with a score less than -0.6 we label "Not Credible", sources with scores above 0.6 we label "Credible", and sources in between are labelled "Unknown". The threshold of 0.6 is inspired from NewsGuards methodology. Table \ref{tbl:newsguard_labels} shows the aggregation of these credibility ratings within each of the communities. 

In addition to credibility ratings, MBFC provides a descriptive label for sites, which often includes the source's political bias across the political spectrum from left to right (using 5 levels). Allsides is another expert-based assessment site, which similarly labels sources with one of 5 levels of bias across the political spectrum from left to right. Finally, BuzzFeed has published a dataset with political leaning of sources, using binary label of either left or right. Using these three political bias assessments, we aggregate a bias score for each source by normalizing each rating from -1 (left) to 1 (right) and adding them. We threshold the scores so that $score \leq -1$ is considered left-leaning, $-1 < score < 1$ is considered center and $1 \leq score$ is considered right-leaning. Table \ref{tab:community_leaning} shows the aggregation the the bias ratings.

\begin{table*}[htbp]
  \centering
  \hspace*{-0.1in}\begin{tabular}{c}
   \includegraphics[width=7in]{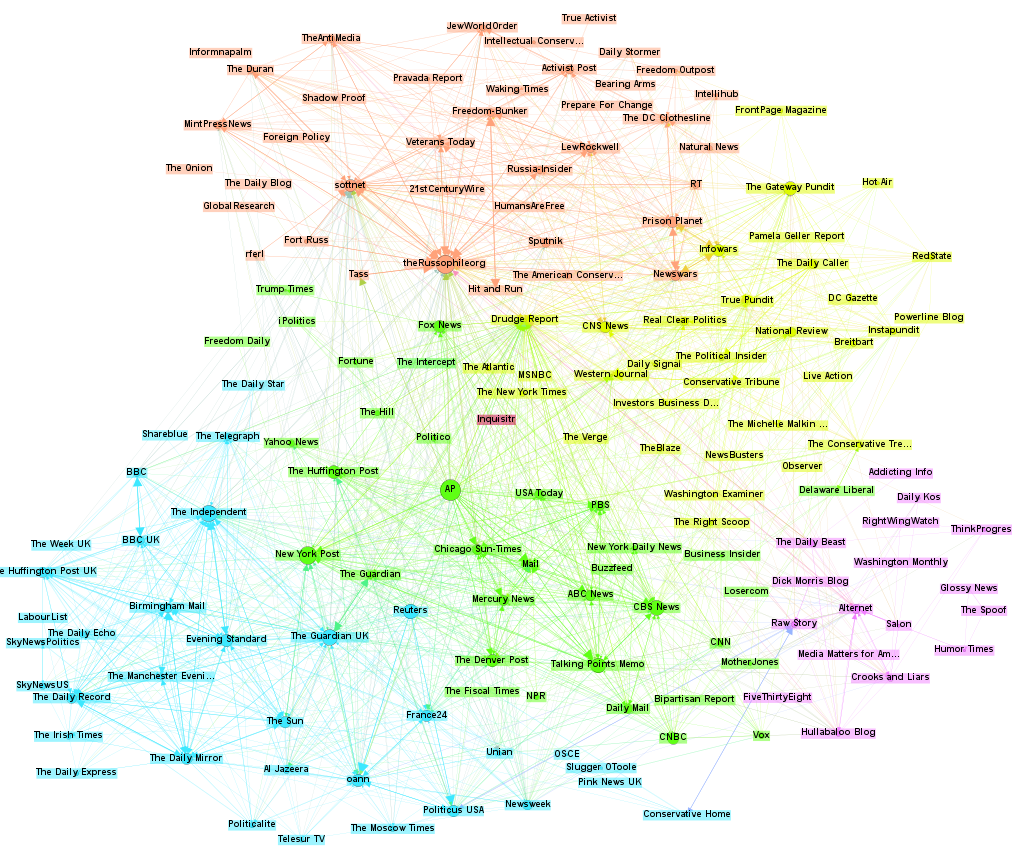}
  \end{tabular}
   \captionof{figure}{Network of news sources, where colors indicated the community membership, size represents outdegree, and arrow direction indicates information flow (e.g. node $n$'s out-degree means how many copy from $n$). The node layout is constructed using Force-Atlas 2 and expansion in Gephi (\url{gephi.org}).}
  \label{nets1}
\end{table*}

\begin{table*}[htbp]
  \centering
    \fontsize{8.8}{11}\selectfont
  \begin{tabular}{cccccc}
  \toprule
   & \textbf{\colorbox{Orange}{ Orange }} & \textbf{\colorbox{Yellow}{ Yellow }} & \textbf{\colorbox{Green}{ Green }} & \textbf{\colorbox{Magenta}{ Magenta }} & \textbf{\colorbox{Cyan}{ Cyan }}\\
   \midrule
   \textbf{N} & 39 & 33 & 35 & 17 & 35 \\
   
   \textbf{Highest Eigenvector Centrality} & The Russophile & Drudge Report & New York Post & Alternet & The Independent\\
      
   \textbf{Most External Outgoing Edges} & Newswars & Infowars & AP & The Daily Beast & Reuters\\
   
   \textbf{Most External Incoming Edges} & Newswars & Drudge Report & The Guardian & Raw Story & The Guardian UK\\
   
  \textbf{ Most Internal Outgoing Edges} & Tass & Conservative Tribune & AP & Alternet & Reuters\\
   
   \textbf{Most Internal Incoming Edges} & The Russophile & Western Journal & Talking Points Memo & Alternet & OANN\\
   \bottomrule
  \end{tabular}
   \caption{Metadata about each community, where ``Most External Outgoing Edges'' shows the node who has the highest number of out directed edges going to nodes in another community, ``Most Internal Outgoing Edges'' shows the node who has the highest number of out directed edges going to nodes in the same community, etc.}
  \label{networkstats}
\end{table*}

\begin{table*}[h]
  \centering
  \fontsize{8.8}{11}\selectfont
  \begin{tabular}{ccccc}
  \toprule
   \textbf{\colorbox{Orange}{ Orange }} & \textbf{\colorbox{Yellow}{ Yellow }} & \textbf{\colorbox{Green}{ Green }} & \textbf{\colorbox{Magenta}{ Magenta }} & \textbf{\colorbox{Cyan}{ Cyan }}\\
   \midrule
   LewRockwell & Infowars  & AP & Salon & The Sun\\
   Prison Planet & Western Journal & PBS & Raw Story & Huffington Post UK\\
   TheAntiMedia  & True Pundit & CBS & Hullabaloo Blog & The Daily Echo\\
   Newswars & CNS News  & Daily Mail& The Daily Beast & The Independent\\
   Russia-Insider & News Busters & The Huffington Post & Crooks and Liars & Birmingham Mail\\
   sottnet & Real Clear Politics & The Guardian& Media Matters & The Daily Record\\
   Mint Press News & Drudge Report & Chicago Sun-Times & Alternet & Evening Standard\\
   The Duran & National Review & The Denver Post & & The Daily Mirror\\
   The Russophile & The Political Insider & ABC & & Manchester Evening News\\
   Activist Post & Investors Business Daily & New York Post & & BBC UK\\
   Freedom-Bunker &  Instapundit & USA Today & & The Guardian UK\\
     & The Daily Caller & Fox News & &\\
     & Daily Signal & Talking Points Memo & &\\
     & The Gateway Pundit & Mercury News & &\\
     \bottomrule
  \end{tabular}
   \caption{Members of the k-core of each community, where the k-core is a maximal subgraph that contains nodes of degree k or more. In this case, we compute the ``main-core" which is the core with the largest degree.}
  \label{kcore}
\end{table*}

\section{Network Analysis Results}~\label{results}
Using our constructed network, we assess the community structure and the traits of the sources within each community to better understand content-sharing. In Figure~\ref{nets1}, we show the network of content sharing, where each color represents a community. In Table~\ref{networkstats}, we show some basic statistics about each community in the network. Our primary results are discussed below. \\

\textbf{Content sharing communities represent distinct parts of the media.} 
In Table~\ref{tbl:newsguard_labels}, we show the breakdown of credibility in each community, in Table~\ref{tab:community_leaning}, we show the breakdown of source political leaning, and in Table~\ref{tab:community_countries}, we show the breakdown of source country in each community. Using these three tables, we can see some clear differences between each community. The green (\colorbox{Green}{G}) community is 89\% U.S. based and 74\% of its sources are credible. This community contains many recognizable mainstream sources, such as AP News, USA Today, NPR, and PBS. The cyan (\colorbox{Cyan}{C}) community contains 63\% U.K. based sources and contains many recognizable mainstream sources such as Reuters, The Independent, and BBC. The magenta (\colorbox{Magenta}{M}) community contains 94\% U.S. based sources that are mostly left leaning. Several of these sources are self-proclaimed liberal blogs, such as Crooks and Liars, RightWingWatch, and Daily Kos. The yellow (\colorbox{Yellow}{Y}) community is 94\% U.S. based and 70\% right leaning. It contains several well-known conspiracy sources, such as Infowars and The Gateway Pundit, as well as many hyper-partisan sources that have published false information in the past, such as The Drudge Report and Breitbart. Lastly, and maybe most interesting, the orange (\colorbox{Orange}{O}) community contains 21\% Russian and 54\% U.S. biased sources. It contains 5\% sources that are marked as not credible and 87\% unknown credibility. Many of the sources are recognizable Russian state-sponsored sources, such as RT and Sputnik, while others are anti-semitic media sources, such as Daily Stormer and JewWorldOrder. In addition to this, there are several right-wing conspiracy sites, such as The D.C. Clothesline, The American Conservative, Natural News, Prison Planet, and Newswars. \\

In the following discussion, we will refer to the communities as:
\begin{description}[itemsep=0pt]
    \item[~\hspace{5mm}\colorbox{Orange}{O}] Russian/conspiracy community
    \item[~\hspace{5mm}\colorbox{Yellow}{Y}] Right-wing/conspiracy community
    \item[~\hspace{5mm}\colorbox{Green}{G}] U.S. mainstream community
    \item[~\hspace{5mm}\colorbox{Magenta}{M}] Left-wing blog community
    \item[~\hspace{5mm}\colorbox{Cyan}{C}] U.K. mainstream  community
\end{description}

Note, there are a few unexpected nodes in the U.S. mainstream community and the right-wing/conspiracy community. Namely, Trump Times in the U.S. mainstream community and The Atlantic, MSNBC, and The New York Times in the right-wing/conspiracy community. There are a few reasons this happens in the network. First, Trump Times, a fairly new right-wing blog, only copies articles from Fox News, which is a right-leaning mainstream news source. Second, unexpectedly, The Atlantic, MSNBC, and The New York Times are all copied multiple times by right-wing sources in the right-wing/conspiracy community. These sources include The Drudge Report, Red State, and Hot Air. The subjects of these articles copied are mostly President Donald Trump's speeches and data privacy. These sources are also heavily copied by members of the green community and members of the magenta community (as expected), but not as much as they are copied by members of the right-wing/conspiracy community. The Atlantic, MSNBC, and The New York Times do not copy from any members of the right-wing/conspiracy community. To further show these sources are peripheral nodes in the network, we compute the k-core of each community, shown in Table~\ref{kcore}. The k-core is the maximal subgraph that contains nodes of degree k or more, which should indicate what sources are most tightly connected in each community. We see that all 4 of these sources do not fall into the k-core of their given communities. \\

\textbf{News-wire services and news aggregators come in all flavors.}
For the most part, each community has its own news-wire-like services, just as was found in~\cite{starbird2018ecosystem}. The U.S. mainstream community has AP News, the U.K. mainstream community has Reuters, and the Russian/conspiracy community has Tass. In the right-wing/conspiracy community, there is not a formal news-wire service as in the other communities, but it seems that Western Journal and Conservative Tribune (which are owned by the same parent company) act as news-wires to much of the community. However, it is also the case that they aggregate many things from both U.S. mainstream media and hyper-partisan right sources. 

Similarly, we see that each community has one or more news aggregators. The U.S. mainstream community has Yahoo News and Mail (\url{www.mail.com}), the right-wing/conspiracy community has The Drudge Report and True Pundit, and the Russian/conspiracy community has The Russophile (a self-proclaimed alternative news aggregator) and sott.net (the self-proclaimed ``leading alternative news site'').\\

\textbf{Hyper-partisan right sources share content more than hyper-partisan left sources.}
It is clear that the right-wing/conspiracy community is much larger than the left-wing blog community (33 sources vs. 17 sources). However, there is a much more balanced set of left and right sources in the full dataset. This finding could be due to conspiracy sources in our dataset being right-wing focused or it could be due to different content sharing behavior among the two groups. When taking a qualitative look at each group of sources, it does seem clear that the left-wing sources write many more long and unique opinion pieces rather than ``breaking news.'' Whereas the right-wing/conspiracy community participates in more breaking news stories, rather than opinion pieces. 

\section{Extending to Partial Content Sharing}~\label{partial}
Now that we have a community framework built using verbatim content sharing and understand its basic characteristics, we extend to partial content sharing by utilizing methods from plagiarism detection. Schleimer et al. creates a method called ``winnowing'', which is a clever combination of hashing and windowing to create fingerprints for text~\cite{schleimer2003winnowing}. These fingerprints refer to a small set of values which can be used to identify pieces of text. The method computes the sequence of hashes of all $k$-grams of characters over a text, for some decided value of $k$. It then runs a window of length $t$ over the hashes and creates a much shorter sequence of minimum hash-values in the windows. This sequence of hash values is the fingerprint used to represent the document. If one compares fingerprints of two documents, the overlapping hashes will (with very high probability) be identical sequences of text. The algorithm can also provide positions of the overlaps in the documents. The authors prove that any sequence of text of length $t$ or more will be detected by the algorithm. Any string of less than $k$ will not be detected. We used $t=25$ and $k=10$. 

After computing fingerprints of all articles, we can very efficiently compare documents and detect overlaps. After detecting overlaps between two documents will analyze the positions of the overlaps and expand to longest ranges of identical text. We combine ranges of matched text that are close to each other (for example, if one article has inserted an additional word into a copied sentence). Finally we perform two thresholds. We only keep segments of copying which are longer than 170 characters and only consider pairs of articles that share segments of a combined size of at least 350 characters. This process leaves us with pairs of articles which share a considerable amount of text. The use of this method in our analysis is describe below.

\begin{figure*}[h]
\begin{floatrow}
\ffigbox{%
\includegraphics[width=3.2in]{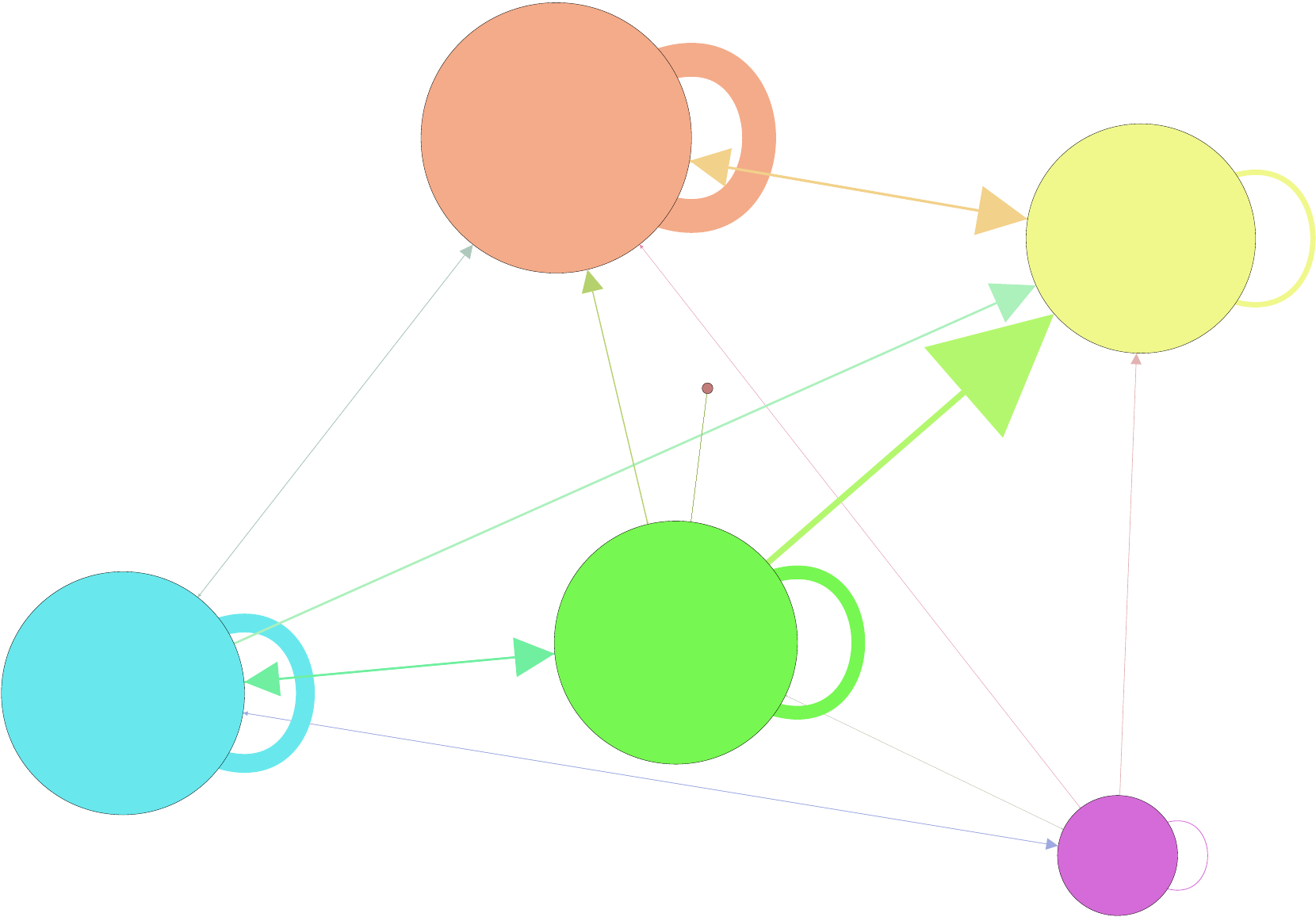}
}{
  \caption{Community level graph where each node is 1 community. The colors correspond to the original community colors, the size of the nodes corresponds to the number of sources in the community, and the size of the edge arrows is the number of copied articles between the communities.}%
  \label{fig:communitylevel}
}
\capbtabbox{%
  \begin{tabular}{cccccc} \toprule
 & \textbf{\colorbox{Orange}{ O }} & \textbf{\colorbox{Yellow}{ Y }} & \textbf{\colorbox{Green}{ G }} & \textbf{\colorbox{Magenta}{ M }} & \textbf{\colorbox{Cyan}{ C }}\\
 \midrule
 \textbf{\colorbox{Orange}{ O }} & 13395 & 1762 & 484 & 102 & 334\\[2mm]
 
 \textbf{\colorbox{Yellow}{ Y }} & & 2133 & 2477 & 225 & 872\\[2mm]
  
 \textbf{\colorbox{Green}{ G }} & &  & 5405 & 66 & 1480\\[2mm]
  
 \textbf{\colorbox{Magenta}{ M }} & & & & 266 & 323\\[2mm]
  
 \textbf{\colorbox{Cyan}{ C }} & & & & & 7448\\[2mm]
 \bottomrule
  \end{tabular}
}{
  \caption{Adjacency matrix where each element in the matrix is the number of article pairs between (or within) each community. Both directions of pairs are counted. }
  \label{tbl:communitylevel}
}
\end{floatrow}
\end{figure*}

\section{Mix-Method Case Studies}
Using the community framework built in Section~\ref{networkbuilding} and the partial content sharing method described in Section~\ref{partial}, we perform a mix-method analysis of content sharing behavior in the network. Specifically, we categorize the prevalent types of article copying that happen in the previously created network. We first look at pairs of articles in the verbatim network to characterize the copying behavior in and out of the communities. Once we have discovered these main types of verbatim copying, we run our partial content matching algorithm to find what other articles, not in the network, spread the given information. We continue to use the community framework described in Section~\ref{results} to describe these findings.

We find four primary behaviors in the data: 
\begin{enumerate}
    \item Echo chambers
    \item Context mixing
    \item Competing narratives
    \item Counter-narratives
\end{enumerate}

\subsection{Echo Chambers}
The most common occurrence in the data were stories that only spread in their given community. This seems fairly clear given our first result in Section~\ref{results}. In Figure~\ref{fig:communitylevel}, we show the community-level network of our original content sharing network, where each node represents a community, and edges represent content sharing between the communities (or within, in the case of self-loops). In Table~\ref{tbl:communitylevel}, we show the weighted adjacency matrix of this network. For the majority of the communities, the number of articles shared within the community (as illustrated by the self-loop weight) is much higher than articles copied from or copied to outside communities. This result should be fairly obvious as the community detection algorithm is looking for the number of links in and between subgraphs. Most of the stories spreading only within a single community are clearly aimed towards the given audience of each community, and are started by a news-wire (or news-wire-like) source such as AP, Reuters, or Tass. Typically, these articles are either location specific (U.S., Russia, Europe, etc.) or political ideology specific (Conservative, Liberal).  

Below we show some representative example stories that stayed within their origin community. We highlight each source name with the community they belong to in Figure~\ref{nets1} for easy referencing. 
\ \\
\begin{itemize}
    \item \colorbox{Orange}{Tass} - Russian fighter jet armed with Kinzhal hypersonic missiles to hold demonstration flights - copied verbatim by \colorbox{Orange}{The Russophile}.
    
    \item \colorbox{Green}{AP} - Nevada could elect first-ever female-majority statehouse - copied verbatim by \colorbox{Green}{Talking Points Memo}, \colorbox{Green}{Mail}, and \colorbox{Green}{CBS News} - copied partially by \colorbox{Green}{The Guardian} and \colorbox{Green}{USA Today}. 
    
    \item \colorbox{Cyan}{France24} - France to set penalties on non-recycled plastic next year - copied verbatim by \colorbox{Cyan}{Telesur TV}
    
    \item \colorbox{Yellow}{The Gateway Pundit} - What is She Wearing Hillary Clinton Looks Like Hell at OzyFest in New York - copied verbatim by \colorbox{Yellow}{The Drudge Report}.
    
    \item \colorbox{Magenta}{Alternet} - Conservatives have gone fully fact-free So how the heck do we even talk to them - copied verbatim by \colorbox{Magenta}{Raw Story}.
\end{itemize}
\ \\

\subsection{Content Mixing}
The second behavior we find in the data is exactly the opposite of echo chambers, namely content mixing. Looking at Figure~\ref{fig:communitylevel}, we can see the yellow (right-wing/conspiracy) community copies many articles from all other communities, particularly the green community (U.S. mainstream). We also see strong connections between the orange (Russian/conspiracy) and yellow (right-wing/conspiracy) communities and between the cyan (U.K. mainstream) and green (U.S. mainstream) communities. 

When further examining the connection between the green and yellow communities, we find that 2038 of the 2477 articles copied from the green community to the yellow community are by The Drudge Report and 160 of these articles are copied by Western Journal. This subtraction leaves us with only 279 article copies by the other sources in the yellow community. Most of the 279 excess articles copied from the mainstream community to the right-wing/conspiracy community are speeches by Donald Trump or reporting about something Donald Trump said, which is ultimately copied verbatim.

While this content mixing between mainstream and right-wing/conspiracy is not as wide-spread as it initially seems, the content mixing that does happen is salient. According to SimilarWeb\footnote{\url{www.similarweb.com (accessed 1/12/2019)}}, The Drudge Report had 138.34M visits in December 2018 and Western Journal had 31.77M visits in December 2018, both with over 90\% of that traffic coming from the United States. According to Alexa\footnote{\url{www.alexa.com} (accessed 1/12/2019)}, both sites were in the top 150 sites visited in the United States. This high readership means that many people are seeing legitimate news articles next to false conspiracy theory articles, potentially creating a warped-view of current events. Furthermore, the employment of so many well-sourced articles may boost the sites' apparent credibility, potentially helping them spread false or misleading information to new readers. 

Content mixing between the right-wing/conspiracy and Russian/conspiracy is much more diverse. Specifically, we see several strong connections between the communities. First, Infowars, Prison Planet, and Newswars form a strongly connected triad that crosses the two communities (with Infowars in right-wing/conspiracy and the other two in Russian/conspiracy). As it turns out, all three sites are ran by the famous conspiracy theorist Alex Jones. More interestingly, all three sources are important in news production in their communities, as all 3 sources show up in the k-core of their communities. Furthermore, all three copy from Russian state-sponsored media such as RT and Sputnik and are copied from by U.S. right-wing sources such as The Gateway Pundit. There are several others sources that cross between the communities, including The D.C. Clothesline, Natural News, The Russophile, and sottnet. The connection between these two communities will be made more even more clear in our next case study. 

We see a similar level of diverse content mixing between the U.K. mainstream and U.S. mainstream communities, but this is to be expected as both are almost completely well-known mainstream sources and content sharing in mainstream media has been studied in depth~\cite{boczkowski2010news}. 

\subsection{Competing Narratives}
Another common finding in the data is competing narratives. Specifically, we often see that news articles are not shared (verbatim or partial) across communities, but the event or topic often is. In other words, completely different articles are written based on the same event and are published in several different news communities, ultimately creating various narratives around a broad event. These competing narratives are often repeatedly shared in the alternative news communities, sometimes multiple times by the same source. This behavior is very similar to the competing narrative behavior surrounding the role of the Syria Civil Defence (White Helmets) shown in \cite{starbird2018ecosystem}. 

A prime example of this in our data is during the Kavanaugh Hearings~\footnote{\url{en.wikipedia.org/wiki/Brett_Kavanaugh_Supreme_Court_nomination#Sexual_assault_allegations}}. On September 22nd, it was announced by Dr. Christine Blasey Ford's lawyers that she would testify in front of the U.S. Senate about accused sexual assault by Supreme Court nominee Brett Kavanaugh. This event had widespread media attention as Kavanaugh's nomination had heavily partisan support and opposition, to which the sexual assault allegations added to. 

In our data set, we can see that this event was not only reported widely, but had many different narratives spreading outside the mainstream media. To illustrate this, we provide some examples of stories spreading around the time of Dr. Ford agreeing to testify. Below we list the articles announcing this event and what news producers either partially or fully used the content. Again, we highlight each source with the community they belong to in Figure~\ref{nets1}. 
\ \\
\begin{itemize}
    \item \colorbox{Cyan}{Reuters} - Kavanaugh accuser agrees to testify in Senate hearing - verbatim copied by \colorbox{Cyan}{OANN}. - partially copied by \colorbox{Green}{NPR}, \colorbox{Green}{CBS News}, \colorbox{Green}{USA Today}, \colorbox{Green}{Fortune}, \colorbox{Green}{Mother Jones}, \colorbox{Cyan}{Politicus USA}, and \colorbox{Orange}{The Russophile}.
    \item \colorbox{Green}{AP} - Kavanaugh accuser commits to hearing - copied verbatim by \colorbox{Green}{Mail}, \colorbox{Yellow}{Drudge Report} - copied partially by \colorbox{Green}{PBS}, \colorbox{Green}{The Denver Post}, \colorbox{Green}{ABC News}, \colorbox{Green}{Mercury News}, and \colorbox{Cyan}{The Independent}.
\end{itemize}
\ \\
Immediately after this event was announced, many competing narratives were spread at the same time in different communities, many of which were fact-checked as false by Politifact\footnote{\url{www.politifact.com}} and Snopes\footnote{\url{www.snopes.com}}. We list examples of these below:
\ \\
\begin{itemize}
    \item \colorbox{Orange}{Natural News} - BOMBSHELL Christine Blasey Fords letter to Sen Dianne Feinstein revealed to be a total FAKE - copied verbatim by \colorbox{Orange}{The D.C. Clothesline} and  \colorbox{Orange}{The Russophile}
    \item \colorbox{Orange}{Natural News} - Kavanaugh accuser Christine Blasey Ford ran mass hypnotic inductions of psychiatric subjects - copied verbatim by \colorbox{Orange}{The Russophile} - copied partially by \colorbox{Orange}{HumansAreFree}.
    \item \colorbox{Yellow}{The Gateway Pundit} - Christine Blasey Fords High School Yearbook Was Scrubbed to Hide Culture of Racism Binge Drinking - copied partially by \colorbox{Orange}{Natural News}, \colorbox{Yellow}{The Right Scoop}, \colorbox{Yellow}{True Pundit}, \colorbox{Orange}{The Russophile}, and \colorbox{Orange}{Sign of the Times} (sottnet).
    \item \colorbox{Yellow}{Hot Air} - Hmmm Ford hires former Clinton Biden adviser as potential hearing sherpa as attorneys bargain for - copied partially by \colorbox{Yellow}{The Gateway Pundit} and \colorbox{Orange}{Sign of the Times} (sottnet)
    \item \colorbox{Yellow}{The Gateway Pundit} - Far Left Activist and Kavanaugh Accuser Christine Blasey Ford Spotted at Anti-Trump March in LA VID - copied partially by The \colorbox{Yellow}{Political Insider} and \colorbox{Yellow}{True Pundit}.
    \item \colorbox{Yellow}{The Gateway Pundit} - Christine Blasey Fords Complete List of Lies and Misrepresentations Related to Judge Kavanaugh - copied partially by \colorbox{Orange}{LewRockwell}, \colorbox{Orange}{The Russophile}.
    \item \colorbox{Yellow}{The Gateway Pundit} - Women Support Judge Brett Kavanaugh Criticize Accuser Dr Christine Blasey Ford - copied partially by \colorbox{Orange}{Prison Planet} and \colorbox{Yellow}{The Daily Caller}.
\end{itemize}
\ \\

Several of these stories where later fact-checked as false. This example illustrates not only the speed of false/hyper-partisan story creation, but the amplification of these stories through content sharing. In this small example, we see 7 different narratives discrediting Dr. Ford before she could even testify. These 7 narratives were all copied multiple times through out both the conspiracy communities. 

We found similar behavior before this event, when the Washington Post revealed Christine Blasey Ford was the accuser. In addition there were many smaller examples of this behavior in the data.

\subsection{Counter-narratives} 
Lastly, we see a more unique case in our data, which we will call (for lack of a better term) counter-narratives. On November 27th, 2018, The Guardian published a story alleging that Paul Manafort, former campaign manager to President Donald Trump, held a secret meeting with Julian Assange, the founder of Wikileaks, inside the Ecuadorian embassy. This story, if true, was considered potentially the biggest story of the year\footnote{\url{vanityfair.com/news/2018/11/the-guardian-paul-manafort-julian-assange}} due to its far reaching implications. However, the story was criticized by other well-reputed sources for relying on anonymous sources, not providing any verifiable details, and being, in general, unbelievable given the high level of surveillance in the area surrounding the embassy\footnote{\url{fair.org/home/misreporting-manafort-a-case-study-in-journalistic-malpractice/}}. The report has been denied by both Manafort and Assange. Additionally, the story was edited by The Guardian multiple times within five hours, weakening the language surrounding the claims. Five weeks after its publication, Glenn Greenwald, a renowned investigative journalist has reported that the story remains unverified\footnote{\url{theintercept.com/2019/01/02/five-weeks-after-the-guardians-viral-blockbuster-assangemanafort-scoop-no-evidence-has-emerged-just-stonewalling/}}. Yet, The Guardian has not retracted the article or demonstrated any further investigation to verify the report. 

This story spread widely in the news ecosystem, spanning all communities. Specifically, we find the following links in our data:

\begin{itemize}
    \item \colorbox{Green}{The Guardian} - Manafort held secret talks with Assange in Ecuadorian embassy sources say - copied verbatim by \colorbox{Yellow}{Drudge Report} and \colorbox{Orange}{The Russophile} - copied partially by \colorbox{Magenta}{Alternet}, \colorbox{Magenta}{Daily Kos}, \colorbox{Magenta}{Crooks and Liars}, \colorbox{Magenta}{Raw Story}, \colorbox{Green}{CNN}, \colorbox{Green}{The Huffington Post}, \colorbox{Green}{Yahoo News}, \colorbox{Green}{Mother Jones}, \colorbox{Green}{Chicago Sun-Times}, \colorbox{Green}{The Hill}, \colorbox{Green}{USA Today}, \colorbox{Green}{Mail} \colorbox{Cyan}{The Independent}, \colorbox{Cyan}{Politicus USA}, \colorbox{Yellow}{Hot Air}, and \colorbox{Yellow}{MSNBC}.
\end{itemize}

This story was also reported by several other mainstream sources, but they did not show up in our partial copy analysis. Our dataset contains verbatim copying of both the original and modified versions of this article.  

While the wide spread of a still unconfirmed story is alarming, as it breaks strongly-held journalist standards, the bigger concern is the aftermath. In the following days, the heavily conspiracy-based communities published and spread the following articles:

\begin{itemize}
    \item \colorbox{Orange}{LewRockwell} - The Assange-Manafort Fabricated Story - copied verbatim by \colorbox{Orange}{The Russophile} and \colorbox{Orange}{Sign of The Times} (sott.net).
    \item \colorbox{Orange}{Natural News} - WOW The Guardian reporters of bogus Manafort-Assange meetings accused of faking stories about WikiL - copied partially by \colorbox{Orange}{Mint Press News}, 
    \item \colorbox{Orange}{Natural News} - The Guardian caught publishing completely fake news - copied partially by \colorbox{Orange}{The Duran}
    \item \colorbox{Yellow}{The Gateway Pundit} - Guardian Stealth Edits Junk Report to Save Their Ass After Assange-Manafort Fiction Crumbles - copied verbatim by \colorbox{Orange}{Prison Planet} and \colorbox{Orange}{Newswars}.
    \item \colorbox{Yellow}{The Gateway Pundit} - IT WAS A HOAX Guardian Report Blows Up Manafort Passport Shows NO UK TRIPS Never Met with Assange
    \item \colorbox{Orange}{Russia-Insider} - Greenwald Goes Ballistic on Politicos Theory That Guardians Assange-Manafort Story Was Planted - copied verbatim by \colorbox{Orange}{Mint Press News}, \colorbox{Orange}{The Russophile}, \colorbox{Orange}{Sign of The Times} (sott.net)
    \item \colorbox{Orange}{21stCenturyWire} - Wikileaks Rips Guardians Manafort-Assange Report by Serial Fabricator Offers Million Dollar Challenge - copied verbatim by \colorbox{Orange}{The Russophile}
    \item \colorbox{Orange}{Sputnik} - As Guardians Manafort-Assange Story Exposed as Fake Ex-CIA Agent Blames Russia
    \item \colorbox{Yellow}{True Pundit} - The Guardian Faceplants As Manaforts Passport Stamps Dont Match Fabricated Assange Story
    \item \colorbox{Orange}{Daily Stormer} - Manafort DID NOT Visit Julian Assange The Guardian Made it All Up
\end{itemize}

At the surface level, these articles seem to be pushing the standard conspiracy narrative that ``the mainstream media is the fake media,'' but in this case their narrative seems to be justified, as the mainstream media is still debating the reliability of The Guardian article. This example illustrates how a small breach of journalistic standards can cause increased uncertainty of what news sources to trust, which ultimately takes power away from the proper news and gives power to conspiracy theorist. Due to the limited attention and information overload of consumers~\cite{qiu2017limited}, this small shift in power may be enough to erode trust. 

\section{Conclusions}
In conclusion, we find that content sharing happens in tightly formed communities, and these communities represent relatively homogeneous portions of the media landscape. We characterize these communities using expert labeling from four independent assessment sites as well as the country the source is based in.  We find many of the same behaviors described in previous studies~\cite{starbird2017examining,horne2018exploration}. Specifically, we find that both news-wire-like services and news aggregation services exist in both alternative and mainstream news communities. We find that alternative news sources repeatedly share content about competing contemporaneous narratives, which can erode trust in the mainstream media as well as cause uncertain surround current events. We also discover mainstream and conspiracy content mixing by several highly read sources in the right-wing community, which can create a false sense of credibility for otherwise not credible sources. In general, our results show that the news is homogeneous within communities and diverse in-between, creating different spirals of sameness. These different spirals have multiple sources working simultaneously to amplify alternative narratives about current events as well as to undermine the credibility of some high quality news outlets.

\section{Limitations and Future Work}
Our goal in this paper was to explore a large-scale view of the news ecosystem, but we recognize our data only covers the English speaking world, and certainly may be incomplete. We do not include local news papers in this study, which also contribute to this environment. Furthermore, we do not fully understand the motives behind much of the content sharing. While some of it may be meant for malicious amplification of fake news, some of it also may be cause by ``useful idiots~\cite{zannettou2018web}.'' These motivations may be hard if not impossible to assess from the outside of these organizations. In addition, the sources used to label sources in the network are incomplete, leaving a lot of data unlabeled, particularly sources from outside the United States. A better understanding of the credibility of these unlabeled sources may help in future analysis. 

In this paper, we did not fully explore the paths created by partial content sharing. For future work, we would like to refine the partial content matching algorithm and continue analysis using it. While the partial sharing cases described in this study were fairly straight-forward, there are many interesting and under-studied cases of real news content being mixed with fake news content. These types of partial copies were not addressed in this paper. 

Lastly, we recognize that the operationalization of sharing may conflate different copying behaviors, such as large quote copying. We attempt to limit this potential conflation by using only verbatim copying in the network construction, which means only fully quote articles, like Presidential speeches, are included in the network. Despite this fix, we would like to find a more sophisticated solution to disambiguating this in future work. 

\small{}
\bibliographystyle{aaai}
\bibliography{references}

\begin{thebibliography}{}

\bibitem[\protect\citeauthoryear{Allcott and
  Gentzkow}{2017}]{allcott2017social}
Allcott, H., and Gentzkow, M.
\newblock 2017.
\newblock Social media and fake news in the 2016 election.
\newblock {\em J. of Economic Perspectives} 31(2):211--36.

\bibitem[\protect\citeauthoryear{Baly \bgroup et al\mbox.\egroup
  }{2018}]{baly2018predicting}
Baly, R.; Karadzhov, G.; Alexandrov, D.; Glass, J.; and Nakov, P.
\newblock 2018.
\newblock Predicting factuality of reporting and bias of news media sources.
\newblock {\em arXiv preprint arXiv:1810.01765}.

\bibitem[\protect\citeauthoryear{Boczkowski}{2010}]{boczkowski2010news}
Boczkowski, P.~J.
\newblock 2010.
\newblock {\em News at work: Imitation in an age of information abundance}.
\newblock University of Chicago Press.

\bibitem[\protect\citeauthoryear{Breed}{1955}]{breed1955newspaper}
Breed, W.
\newblock 1955.
\newblock Newspaper ‘opinion leaders’ and processes of standardization.
\newblock {\em Journalism Quarterly} 32(3):277--328.

\bibitem[\protect\citeauthoryear{Chakraborty \bgroup et al\mbox.\egroup
  }{2016}]{chakraborty2016stop}
Chakraborty, A.; Paranjape, B.; Kakarla, S.; and Ganguly, N.
\newblock 2016.
\newblock Stop clickbait: Detecting and preventing clickbaits in online news
  media.
\newblock In {\em ASONAM},  9--16.
\newblock IEEE.

\bibitem[\protect\citeauthoryear{Glasser}{1992}]{glasser1992professionalism}
Glasser, T.~L.
\newblock 1992.
\newblock Professionalism and the derision of diversity: The case of the
  education of journalists.
\newblock {\em Journal of communication} 42(2):131--140.

\bibitem[\protect\citeauthoryear{Graber}{1971}]{graber1971press}
Graber, D.
\newblock 1971.
\newblock The press as opinion resource during the 1968 presidential campaign.
\newblock {\em Public opinion quarterly} 35(2):168--182.

\bibitem[\protect\citeauthoryear{Horne and Adal{\i}}{2017}]{horne2017just}
Horne, B.~D., and Adal{\i}, S.
\newblock 2017.
\newblock This just in: Fake news packs a lot in title, uses simpler,
  repetitive content in text body, more similar to satire than real news.
\newblock In {\em ICWSM NECO Workshop}.

\bibitem[\protect\citeauthoryear{Horne and
  Adal{\i}}{2018}]{horne2018exploration}
Horne, B.~D., and Adal{\i}, S.
\newblock 2018.
\newblock An exploration of verbatim content republishing by news producers.
\newblock {\em arXiv preprint arXiv:1805.05939}.

\bibitem[\protect\citeauthoryear{Horne \bgroup et al\mbox.\egroup
  }{2018}]{horne2018accessing}
Horne, B.~D.; Dron, W.; Khedr, S.; and Adal{\i}, S.
\newblock 2018.
\newblock Assessing the news landscape: A multi-module toolkit for evaluating
  the credibility of news.
\newblock In {\em WWW Companion}.

\bibitem[\protect\citeauthoryear{Klinenberg}{2005}]{klinenberg2005convergence}
Klinenberg, E.
\newblock 2005.
\newblock Convergence: News production in a digital age.
\newblock {\em The Annals of the American Academy of Political and Social
  Science} 597(1):48--64.

\bibitem[\protect\citeauthoryear{Leicht and Newman}{2008}]{leicht2008a}
Leicht, E.~A., and Newman, M. E.~J.
\newblock 2008.
\newblock Community structure in directed networks.
\newblock {\em Physical Review Letters} 100(11):118703.

\bibitem[\protect\citeauthoryear{Mele \bgroup et al\mbox.\egroup
  }{2017}]{mele2017combating}
Mele, N.; Lazer, D.; Baum, M.; Grinberg, N.; Friedland, L.; Joseph, K.; Hobbs,
  W.; and Mattsson, C.
\newblock 2017.
\newblock Combating fake news: An agenda for research and action.

\bibitem[\protect\citeauthoryear{Mitchelstein and
  Boczkowski}{2010}]{mitchelstein2010online}
Mitchelstein, E., and Boczkowski, P.~J.
\newblock 2010.
\newblock Online news consumption research: An assessment of past work and an
  agenda for the future.
\newblock {\em New Media \& Society} 12(7):1085--1102.

\bibitem[\protect\citeauthoryear{Noelle-Neumann and
  Mathes}{1987}]{noelle1987theevent}
Noelle-Neumann, E., and Mathes, R.
\newblock 1987.
\newblock Theevent as event'and theevent as news': The significance
  ofconsonance'for media effects research.
\newblock {\em European Journal of Communication} 2(4):391--414.

\bibitem[\protect\citeauthoryear{Popat \bgroup et al\mbox.\egroup
  }{2016}]{popat2016credibility}
Popat, K.; Mukherjee, S.; Str{\"o}tgen, J.; and Weikum, G.
\newblock 2016.
\newblock Credibility assessment of textual claims on the web.
\newblock In {\em CIKM},  2173--2178.
\newblock ACM.

\bibitem[\protect\citeauthoryear{Potthast \bgroup et al\mbox.\egroup
  }{2017}]{potthast2017stylometric}
Potthast, M.; Kiesel, J.; Reinartz, K.; Bevendorff, J.; and Stein, B.
\newblock 2017.
\newblock A stylometric inquiry into hyperpartisan and fake news.
\newblock {\em arXiv preprint arXiv:1702.05638}.

\bibitem[\protect\citeauthoryear{Qiu \bgroup et al\mbox.\egroup
  }{2017}]{qiu2017limited}
Qiu, X.; Oliveira, D.~F.; Shirazi, A.~S.; Flammini, A.; and Menczer, F.
\newblock 2017.
\newblock Limited individual attention and online virality of low-quality
  information.
\newblock {\em Nature Human Behaviour} 1(7):0132.

\bibitem[\protect\citeauthoryear{Reese, Vos, and
  Shoemaker}{2009}]{reese2009journalists}
Reese, S.~D.; Vos, T.~P.; and Shoemaker, P.~J.
\newblock 2009.
\newblock Journalists as gatekeepers.
\newblock In {\em The handbook of journalism studies}. Routledge.
\newblock  93--107.

\bibitem[\protect\citeauthoryear{Schleimer, Wilkerson, and
  Aiken}{2003}]{schleimer2003winnowing}
Schleimer, S.; Wilkerson, D.~S.; and Aiken, A.
\newblock 2003.
\newblock Winnowing: local algorithms for document fingerprinting.
\newblock In {\em Proceedings of the 2003 ACM SIGMOD international conference
  on Management of data},  76--85.
\newblock ACM.

\bibitem[\protect\citeauthoryear{Shao \bgroup et al\mbox.\egroup
  }{2017}]{shao2017spread}
Shao, C.; Ciampaglia, G.~L.; Varol, O.; Flammini, A.; and Menczer, F.
\newblock 2017.
\newblock The spread of fake news by social bots.
\newblock {\em arXiv preprint arXiv:1707.07592}.

\bibitem[\protect\citeauthoryear{Shoemaker and
  Reese}{2013}]{shoemaker2013mediating}
Shoemaker, P.~J., and Reese, S.~D.
\newblock 2013.
\newblock {\em Mediating the message in the 21st century: A media sociology
  perspective}.
\newblock Routledge.

\bibitem[\protect\citeauthoryear{Singhania, Fernandez, and
  Rao}{2017}]{singhania20173han}
Singhania, S.; Fernandez, N.; and Rao, S.
\newblock 2017.
\newblock 3han: A deep neural network for fake news detection.
\newblock In {\em International Conference on Neural Information Processing},
  572--581.
\newblock Springer.

\bibitem[\protect\citeauthoryear{Starbird \bgroup et al\mbox.\egroup
  }{2018}]{starbird2018ecosystem}
Starbird, K.; Arif, A.; Wilson, T.; Van~Koevering, K.; Yefimova, K.; and
  Scarnecchia, D.
\newblock 2018.
\newblock Ecosystem or echo-system? exploring content sharing across
  alternative media domains.

\bibitem[\protect\citeauthoryear{Starbird}{2017}]{starbird2017examining}
Starbird, K.
\newblock 2017.
\newblock Examining the alternative media ecosystem through the production of
  alternative narratives of mass shooting events on twitter.
\newblock In {\em ICWSM},  230--239.

\bibitem[\protect\citeauthoryear{Zannettou \bgroup et al\mbox.\egroup
  }{2018}]{zannettou2018web}
Zannettou, S.; Sirivianos, M.; Blackburn, J.; and Kourtellis, N.
\newblock 2018.
\newblock The web of false information: Rumors, fake news, hoaxes, clickbait,
  and various other shenanigans.
\newblock {\em arXiv preprint arXiv:1804.03461}.

\end{thebibliography}

\end{document}